# Title: Observation of Long-Lived Interlayer Excitons in Monolayer MoSe$_2$-WSe$_2$ Heterostructures

**Authors**: Pasqual Rivera[1], John R. Schaibley[1], Aaron M. Jones[1], Jason S. Ross[2], Sanfeng Wu[1], Grant Aivazian[1], Philip Klement[1] , Nirmal J. Ghimire[3,4], Jiaqiang Yan[4,5], D. G. Mandrus[3,4,5], Wang Yao[6], Xiaodong Xu[1,2*]

Affiliations:
[1]Department of Physics, University of Washington, Seattle, Washington 98195, USA
[2]Department of Materials Science and Engineering, University of Washington, Seattle, Washington 98195, USA
[3]Department of Physics and Astronomy, University of Tennessee, Knoxville, Tennessee 37996, USA
[4]Materials Science and Technology Division, Oak Ridge National Laboratory, Oak Ridge, Tennessee, 37831, USA
[5]Department of Materials Science and Engineering, University of Tennessee, Knoxville, Tennessee, 37996, USA
[6]Department of Physics and Center of Theoretical and Computational Physics, University of Hong Kong, Hong Kong, China

*Correspondence to:  xuxd@uw.edu

**Abstract: Two-dimensional (2D) materials, such as graphene[1], boron nitride[2], and transition metal dichalcogenides (TMDs)[3-5], have sparked wide interest in both device physics and technological applications at the atomic monolayer limit. These 2D monolayers can be stacked together with precise control to form novel van der Waals heterostructures for new functionalities[2,6-9]. One highly coveted but yet to be realized heterostructure is that of differing monolayer TMDs with type II band alignment[10-12]. Their application potential hinges on the fabrication, understanding, and control of bonded monolayers, with bound electrons and holes localized in individual monolayers, i.e. interlayer excitons. Here, we report the first observation of interlayer excitons in monolayer MoSe$_2$-WSe$_2$ heterostructures by both photoluminescence and photoluminescence excitation spectroscopy. The energy and luminescence intensity of interlayer excitons are highly tunable by an applied vertical gate voltage, implying electrical control of the heterojunction band-alignment. Using time resolved photoluminescence, we find that the interlayer exciton is long-lived with a lifetime of about 1.8 ns, an order of magnitude longer than intralayer excitons[13-16]. Our work demonstrates the ability to optically pump interlayer electric polarization and provokes the immediate exploration of interlayer excitons for condensation phenomena, as well as new applications in 2D light-emitting diodes, lasers, and photovoltaic devices.**

**Main Text**

The recently developed ability to vertically assemble different 2D materials heralds a new realm of device physics based on van der Waals heterostructures[7]. The most successful example is the vertical integration of graphene on boron nitride. Such novel heterostructures not only dramatically enhance graphene's electronic properties[2], but also give rise to super-lattice structures demonstrating exotic physical phenomena[6,8-9]. A fascinating counterpart to gapless graphene is a class of monolayer direct bandgap semiconductors, namely transition metal dichalcogenides (TMDs)[3-5]. Due to the large binding energy in these 2D semiconductors, excitons dominate the optical response, exhibiting strong light-matter interactions which are electrically tunable[17-18]. The discovery of excitonic valley physics[19-23] and strongly coupled spin and pseudospin physics [24-25] in 2D TMDs opens up new possibilities for device concepts not possible in other material systems.

Monolayer semiconductor TMDs have the chemical formula $MX_2$ where the M is tungsten (W) or molybdenum (Mo), and the X is sulfur (S) or selenium (Se). Although these TMDs share the same crystalline structure, their physical properties, such as bandgap, exciton resonance, and spin-orbit coupling strengths, can vary significantly. Therefore, an intriguing possibility is to stack different TMD monolayers on top of one another to form 2D heterostructures. Recent theoretical works have predicted type II band-alignment of such monolayer TMD heterojunctions[10-12], where the conduction band minimum and the valence band maximum are located in different layers. Remarkably, the Coulomb binding energy in 2D TMDs is much stronger than in conventional semiconductors, making it possible to realize interlayer excitonic states in van der Waals bound hetero-bilayers with type II band-alignment. These interlayer excitons are composed of bound electrons and holes that are localized in different layers. Such interlayer excitons have been intensely pursued in bilayer graphene for possible exciton condensation[26], but direct optical observation demonstrating the existence of such excitons is challenging due to the lack of a sizable bandgap in graphene. Monolayer TMDs with bandgaps in the visible range provide the opportunity to optically pump interlayer excitons, which are directly observable through photoluminescence (PL) measurements.

Heterostructures are prepared by mechanical exfoliation of monolayer $MoSe_2$ onto 300 nm thick $SiO_2$ on a heavily doped Si substrate, and $WSe_2$ onto PMMA on a thin layer of PVA atop a

Si substrate. The WSe$_2$ monolayer is then transferred on top of the MoSe$_2$ after which the polymer resist is dissolved in acetone. A top down view of the idealized vertical MoSe$_2$-WSe$_2$ heterostructure is depicted in Fig. 1a. We have fabricated six devices which all show similar results. The data presented here are from two independent MoSe$_2$-WSe$_2$ heterostructures, labelled Device 1 and Device 2. Figure 1b shows an optical micrograph of Device 1. The staggered stacking provides areas of individual monolayers, as well as a large area of the vertically stacked heterostructure.

We characterized the MoSe$_2$-WSe$_2$ heterostructure using photoluminescence (PL) measurements with 660 nm laser excitation (1.88 eV). Inspection of the PL from the heterostructure at room temperature reveals three dominant spectral features (Fig. 1c from Device 1). The emission at 1.65 eV and 1.57 eV corresponds to the excitonic states from monolayer WSe$_2$ and MoSe$_2$[17,23,27], respectively. PL from the heterostructure region reveals a distinct spectral feature at 1.35 eV, which we attribute to interlayer exciton emission. We then perform low temperature optical measurements to study the structure of these intralayer and interlayer excitons.

Figure 1d shows PL spectra from individual monolayer WSe$_2$ (top), MoSe$_2$ (bottom), and the heterostructure area (middle) at a temperature of 20 K. At low temperature, the intralayer neutral ($X_M^o$) and charged ($X_M^-$) excitons are easily resolved[17,23], where M labels either W or Mo. Comparison of the three spectra shows that both intralayer $X_M^o$ and $X_M^-$ exist in the heterostructure with emission at the same energy as from isolated monolayers. This clearly demonstrates the preservation of intralayer excitons in the heterostructure region. The new spectral feature ($X_I$) is more pronounced at low temperature, having an intensity comparable to the intralayer excitons.

The PL spectra imply type II band-alignment for the WSe$_2$-MoSe$_2$ heterojunction, consistent with several independent theoretical predictions[10-12]. This four-level band structure, as shown in Fig. 2a, leads to electron (hole) transfer from WSe$_2$ (MoSe$_2$) to MoSe$_2$ (WSe$_2$), where the lowest energy state resides. The strong Coulomb attraction between electrons in the MoSe$_2$ and holes in the WSe$_2$ (Fig. 2a) then gives rise to the formation of interlayer excitons, evidenced by the new excitonic feature $X_I$ centered at 1.40 eV. The increase of interlayer exciton PL with decreasing temperature (Figure S1, Supplementary Materials) is consistent with the fact that it is the lowest energy bright excitonic state.

From the intralayer and interlayer exciton spectral positions, we can infer the band offset between individual monolayers in the observed type II heterojunction. The energy difference between $X_W$ and $X_I$ at room temperature is 310 meV. Considering the smaller binding energy of interlayer than intralayer excitons, this sets a lower bound on the conduction band offset. The energy difference $(X_{Mo} - X_I) = 230$ meV then provides a lower bound on the valence band offset. The observation of bright interlayer excitons in monolayer semiconducting heterostructures is of central importance and the rest of this paper will focus on their physical properties.

Our PL measurements show that the intralayer exciton emission at the heterostructure region is quenched compared to isolated monolayers. This observation implies that interlayer charge transfer is fast compared to the intralayer exciton recombination rate. Importantly, there is a distinct difference in the quenching ratio at room temperature and low temperature. At room temperature, the PL of intralayer excitons $X_{Mo}$ and $X_W$ is quenched by at least an order of magnitude (Figure S2), while at low temperature the intralayer exciton PL is only slightly quenched (Fig. 1d). This implies that the interlayer carrier hopping rate strongly depends on the temperature and is reduced at lower temperature.

Moreover, at low temperature (20 K), we find the spectrally integrated exciton PL intensity is conserved between isolated monolayers and the heterojunction, while at room temperature the integrated PL intensity from the heterojunction is an order of magnitude smaller than the summation of isolated monolayers. This observation implies that at low temperature the quenched population of intralayer excitons is transferred to form interlayer excitons, which then relax mainly via radiative recombination. In contrast, at room temperature, a non-radiative relaxation channel dominates the interlayer exciton relaxation, giving rise to the quenching of the spectrally integrated heterojunction PL intensity.

To investigate the coupling between the interlayer and intralayer excitons, we perform PL excitation spectroscopy (PLE). A narrow bandwidth (<50 kHz) tunable laser is swept in energy from 1.6 eV to 1.75 eV (across the resonance of intralayer excitons) while monitoring interlayer exciton PL. Figure 2b shows an intensity plot of $X_I$ emission as a function of photo-excitation energy from Device 2. We clearly observe interlayer emission enhancement when the excitation energy is resonant with intralayer exciton states (Fig. 2c). We attribute the PLE resonances to the excitation of intralayer excitons, which then relax through fast interlayer charge transfer to form the energetically favorable interlayer excitons (Fig. 2a).

Furthermore, the emission energy of the interlayer exciton is highly electrically tunable. Figure 3a shows the contact geometry of Device 2. An indium contact is fabricated on the $WSe_2$ layer which is located on top of the $MoSe_2$. The heterostructure is on $SiO_2$, which provides insulation from the heavily doped silicon backgate. We electrostatically dope the heterostructure by grounding the indium contact and applying a voltage to the backgate ($V_g$). Figure 3c shows the interlayer exciton PL intensity as a function of applied gate from +100 to −100 V. Over this range we observe a nearly 3-fold increase in intensity while the peak center blue shifts by approximately 45 meV.

The above observation signifies direct electrical control of the heterojunction band alignment by an externally applied perpendicular electric field. The interlayer exciton, comprised of an electron in the conduction band of $MoSe_2$ and a hole in the $WSe_2$ valence band, has a permanent dipole pointing from $MoSe_2$ to $WSe_2$. Thus, the electric polarization between layers of the heterostructure depends on the exciton density. As shown in Figure 3b-c, at negative $V_g$ the electric field reduces the relative band-offset between the $MoSe_2$ and $WSe_2$ and increases the energy separation between the $MoSe_2$ conduction band and the $WSe_2$ valence band, resulting in the observed blue shift of interlayer exciton PL. At the same time, with the decreasing conduction and valence band offsets between layers, the interlayer charge hopping rate increases, giving rise to an increased intensity of the interlayer exciton PL.

Power dependent measurements reveal interesting physical properties of the interlayer exciton, including a repulsive exciton-exciton interaction, a much longer lifetime compared to intralayer excitons, as well as evidence for the spin splitting of the $MoSe_2$ conduction band. Figure 4a shows the normalized interlayer exciton PLE spectrum as a function of laser power with excitation energy in resonance with $X_W^o$ (1.722 eV). The interlayer exciton is composed of a doublet where both peaks blue shift with increasing power. We attribute the blue shift of the doublet to the repulsive interaction between the dipole-aligned interlayer excitons (*c.f.* Fig. 3a), similar to the spatially indirect excitons in GaAs double quantum wells[30]. Increasing excitation power leads to higher densities and hence an increased energy of interlayer excitons.

We attribute the doublet feature of the interlayer exciton to the spin-splitting of the $MoSe_2$ conduction band[28-29] (Fig. 4b), which is supported by the evolution of the relative strength of the two peaks with increasing excitation power, as shown in Fig. 4a (similar results in Device 1 with 1.88 eV excitation shown in Fig. S3). The emission intensity of the lower energy peak of the doublet dominates at low laser power, while at high power the two peaks have comparable intensities. At low power, the lowest energy configuration of interlayer excitons with the electron in the lower spin-split band of $MoSe_2$, is populated first. Due to phase space filling effects, the interlayer exciton configuration with the electron in the higher spin-split band starts to be filled at higher laser power. Consequently, the higher energy peak of the doublet becomes more prominent at higher excitation powers, as highlighted by the bi-Lorentzian fits at 5 and 100 µW in Figure 4a. The extracted energy difference between the interlayer exciton doublet is about 25 meV, which agrees with $MoSe_2$ conduction band splitting predicted by first principle calculations[28-29].

Figure 4c shows the integrated interlayer exciton emission intensity in Fig. 4a as a function of laser power, normalized for excitation power and integration time. An important observation is that the interlayer exciton peak saturates at low intensity (< 0.5 W/cm$^2$), while independent measurements show that intralayer excitons do not saturate up to the highest applied excitation intensity (>100 W/cm$^2$). This low power saturation implies that the interlayer exciton has a much longer lifetime than that of intralayer excitons. In the heterostructure, the lifetime of the intralayer exciton is substantially reduced by the fast interlayer charge hopping, while the lifetime of the interlayer exciton is long because it is the lowest energy configuration and its spatially indirect nature leads to a reduced optical dipole moment. This long lifetime is confirmed by time resolved PL which directly measures the lifetime of the interlayer exciton, as shown in Figure 4d. A fit to a single exponential decay yields an interlayer exciton lifetime of about 1.8 ns. This time scale is much longer than the intralayer exciton lifetime, which is on the order of tens of ps[13-16].

In summary, our results demonstrate the feasibility of band engineering by creating monolayer TMD van der Waals bound heterostructures. Photoluminescence spectroscopy of interlayer and intralayer excitons with electrical control reveals the type II band-alignment of the heterojunction, crucial for many optoelectronic applications. The interlayer exciton in these heterostructures is an analog of the spatially indirect exciton in GaAs double well structures which have been intensely studied for exciton BEC phenomena[26,30]. Here, we have identified spatially indirect interlayer excitons displaying extended lifetimes and repulsive interactions, both key ingredients for the realization of exciton BEC. The long-lived interlayer exciton may also lead to population inversion, a critical step toward realizing 2D heterostructure lasers.

During the final preparation of the paper, we became aware of relevant work studying the photovoltaic properties[31-33] and interlayer coupling[34] of heterostructures.

## METHODS

Monolayers of $MoSe_2$ are mechanically exfoliated onto 300 nm $SiO_2$ on heavily doped Si wafers and monolayers of $WSe_2$ onto a layer of PMMA atop PVA on Si. Both monolayers are identified with an optical microscope and confirmed by their photoluminescence spectra. After transferring the $WSe_2$, the PMMA is dissolved in acetone for ~30 min and then rinsed with IPA. Low temperature measurements are conducted in a temperature controlled Janis cold finger cryostat (sample in vacuum) with excitation beam diameter of ~1 μm. PL is spectrally filtered through a 0.5 m Andor Shamrock monochromator and detected on an Andor Newton CCD. For PLE measurements, a continuous wave Ti:Sapphire laser (MSquared - SolsTiS) is used for excitation and filtered from the PL signal using an 815 nm long pass optical filter (Semrock). Electrostatic doping is accomplished with an indium drain contact deposited onto the monolayer $WSe_2$ region of Device 2 and using the heavily doped Si as a tunable backgate. For interlayer lifetime measurements, we excite the sample with a <200 fs pulsed Ti:Sapphire laser (Coherent - MIRA). Interlayer PL is spectrally filtered through a 0.5 m monochromator (Princeton Acton 2500), and detected with a fast time correlated single photon counting system composed of a fast (<30 ps FWHM) single photon avalanche detector (Micro Photon Devices-PDM series) and a picosecond event timer (PicoQuant- PicoHarp 300).

## ACKNOWLEDGEMENTS

This work is mainly supported by the US DoE, BES, Materials Sciences and Engineering Division (DE-SC0008145). NG, JY, DM are supported by US DoE, BES, Materials Sciences and Engineering Division. WY is supported by the Research Grant Council of Hong Kong (HKU705513P, HKU8/CRF/11G), and the Croucher Foundation under the Croucher Innovation

Award. XX thanks the support of Cottrell Scholar Award. PR thanks the UW GO-MAP program for their support. AMJ partially supported by the NSF (DGE-0718124). Device fabrication was performed at the Washington Nanofabrication Facility and NSF-funded Nanotech User Facility.

## AUTHOR CONTRIBUTIONS

XX conceived the experiments. PR and PK fabricated the devices, assisted by JSR. PR performed the measurements, assisted by JRS, AMJ, JSR, SW and GA. PR and XX performed data analysis, with input from WY. NG, JY and DGM synthesized and characterized the bulk $WSe_2$ crystals. XX, PR, JRS, and WY wrote the paper. All authors discussed the results.

## AUTHOR INFORMATION

The authors declare no competing financial interests. Correspondence and requests for materials should be addressed to XX.

## REFERENCES

1   Novoselov, K. S., Geim, A. K., Morozov, S. V., Jiang, D., Zhang, Y., Dubonos, S. V., Grigorieva, I. V. & Firsov, A. A. Electric Field Effect in Atomically Thin Carbon Films. *Science* **306**, 666-669, doi:10.1126/science.1102896 (2004).
2   Dean, C. R., Young, A. F., MericI, LeeC, WangL, SorgenfreiS, WatanabeK, TaniguchiT, KimP, Shepard, K. L. & HoneJ. Boron nitride substrates for high-quality graphene electronics. *Nat Nano* **5**, 722-726 (2010).
3   Novoselov, K. S., Jiang, D., Schedin, F., Booth, T. J., Khotkevich, V. V., Morozov, S. V. & Geim, A. K. Two-dimensional atomic crystals. *PNAS* **102**, 10451-10453 (2005).
4   Mak, K. F., Lee, C., Hone, J., Shan, J. & Heinz, T. F. Atomically Thin $MoS_2$: A New Direct-Gap Semiconductor. *Physical Review Letters* **105**, 136805 (2010).
5   Splendiani, A., Sun, L., Zhang, Y., Li, T., Kim, J., Chim, C.-Y., Galli, G. & Wang, F. Emerging photoluminescence in monolayer MoS2. *Nano Lett.* **10**, 1271-1275 (2010).
6   Dean, C. R., Wang, L., Maher, P., Forsythe, C., Ghahari, F., Gao, Y., Katoch, J., Ishigami, M., Moon, P., Koshino, M., Taniguchi, T., Watanabe, K., Shepard, K. L., Hone, J. & Kim, P. Hofstadter/'s butterfly and the fractal quantum Hall effect in moire superlattices. *Nature* **497**, 598-602 (2013).
7   Geim, A. K. & Grigorieva, I. V. Van der Waals heterostructures. *Nature* **499**, 419-425, doi:10.1038/nature12385 (2013).
8   Ponomarenko, L. A., Gorbachev, R. V., Yu, G. L., Elias, D. C., Jalil, R., Patel, A. A., Mishchenko, A., Mayorov, A. S., Woods, C. R., Wallbank, J. R., Mucha-Kruczynski, M., Piot, B. A., Potemski, M., Grigorieva, I. V., Novoselov, K. S., Guinea, F., Fal/'ko, V. I. & Geim, A. K. Cloning of Dirac fermions in graphene superlattices. *Nature* **497**, 594-597 (2013).
9   Hunt, B., Sanchez-Yamagishi, J. D., Young, A. F., Yankowitz, M., LeRoy, B. J., Watanabe, K., Taniguchi, T., Moon, P., Koshino, M., Jarillo-Herrero, P. & Ashoori, R. C. Massive Dirac Fermions and Hofstadter Butterfly in a van der Waals Heterostructure. *Science* **340**, 1427-1430 (2013).

10  Kang, J., Tongay, S., Zhou, J., Li, J. & Wu, J. Band offsets and heterostructures of two-dimensional semiconductors. *Applied Physics Letters* **102**, 012111-012114 (2013).
11  Kośmider, K. & Fernández-Rossier, J. Electronic properties of the $MoS_2$-$WS_2$ heterojunction. *Physical Review B* **87**, 075451 (2013).
12  Terrones, H., López-Urías, F. & Terrones, M. Novel hetero-layered materials with tunable direct band gaps by sandwiching different metal disulfides and diselenides. *Sci. Rep.* **3** (2013).
13  Wang, G., Bouet, L., Lagarde, D., Vidal, M., Balocchi, A., Amand, T., Marie, X. & Urbaszek, B. Valley dynamics probed through charged and neutral exciton emission in monolayer WSe2. *arXiv:1402.6009* (2014).
14  Lagarde, D., Bouet, L., Marie, X., Zhu, C. R., Liu, B. L., Amand, T., Tan, P. H. & Urbaszek, B. Carrier and Polarization Dynamics in Monolayer $MoS_2$. *Physical Review Letters* **112**, 047401 (2014).
15  Mai, C., Barrette, A., Yu, Y., Semenov, Y. G., Kim, K. W., Cao, L. & Gundogdu, K. Many-Body Effects in Valleytronics: Direct Measurement of Valley Lifetimes in Single-Layer MoS2. *Nano Lett.* **14**, 202-206, doi:10.1021/nl403742j (2013).
16  Shi, H., Yan, R., Bertolazzi, S., Brivio, J., Gao, B., Kis, A., Jena, D., Xing, H. G. & Huang, L. Exciton Dynamics in Suspended Monolayer and Few-Layer MoS2 2D Crystals. *ACS Nano* **7**, 1072-1080 (2012).
17  Ross, J. S., Wu, S., Yu, H., Ghimire, N. J., Jones, A. M., Aivazian, G., Yan, J., Mandrus, D. G., Xiao, D., Yao, W. & Xu, X. Electrical control of neutral and charged excitons in a monolayer semiconductor. *Nat Commun* **4**, 1474, doi:10.1038/ncomms2498 (2013).
18  Mak, K. F., He, K., Lee, C., Lee, G. H., Hone, J., Heinz, T. F. & Shan, J. Tightly bound trions in monolayer MoS2. *Nat Mater* **12**, 207-211 (2013).
19  Xiao, D., Liu, G.-B., Feng, W., Xu, X. & Yao, W. Coupled Spin and Valley Physics in Monolayers of $MoS_2$ and Other Group-VI Dichalcogenides. *Physical Review Letters* **108**, 196802 (2012).
20  Cao, T., Wang, G., Han, W., Ye, H., Zhu, C., Shi, J., Niu, Q., Tan, P., Wang, E., Liu, B. & Feng, J. Valley-selective circular dichroism of monolayer molybdenum disulphide. *Nat Commun* **3**, 887 (2012).
21  Zeng, H., Dai, J., Yao, W., Xiao, D. & Cui, X. Valley polarization in MoS2 monolayers by optical pumping. *Nat Nano* **7**, 490-493, doi:10.1038/nnano.2012.95 (2012).
22  Mak, K. F., He, K., Shan, J. & Heinz, T. F. Control of valley polarization in monolayer $MoS_2$ by optical helicity. *Nat Nano* **7**, 494-498 (2012).
23  Jones, A. M., Yu, H., Ghimire, N. J., Wu, S., Aivazian, G., Ross, J. S., Zhao, B., Yan, J., Mandrus, D. G., Xiao, D., Yao, W. & Xu, X. Optical generation of excitonic valley coherence in monolayer WSe2. *Nat Nano* **8**, 634-638 (2013).
24  Gong, Z., Liu, G.-B., Yu, H., Xiao, D., Cui, X., Xu, X. & Yao, W. Magnetoelectric effects and valley-controlled spin quantum gates in transition metal dichalcogenide bilayers. *Nat Commun* **4**, doi:10.1038/ncomms3053 (2013).
25  Jones, A. M., Yu, H., Ross, J. S., Klement, P., Ghimire, N. J., Yan, J., Mandrus, D. G., Yao, W. & Xu, X. Spin-layer locking effects in optical orientation of exciton spin in bilayer WSe2. *Nat Phys* **10**, 130-134 (2014).
26  Su, J.-J. & MacDonald, A. H. How to make a bilayer exciton condensate flow. *Nat Phys* **4**, 799-802 (2008).
27  Zhao, W., Ghorannevis, Z., Chu, L., Toh, M., Kloc, C., Tan, P.-H. & Eda, G. Evolution of Electronic Structure in Atomically Thin Sheets of WS2 and WSe2. *ACS Nano* **7**, 791-797 (2013).
28  Cheiwchanchamnangij, T. & Lambrecht, W. R. L. Quasiparticle band structure calculation of monolayer, bilayer, and bulk $MoS_2$. *Physical Review B* **85**, 205302 (2012).
29  Kormányos, A., Zólyomi, V., Drummond, N. D., Rakyta, P., Burkard, G. & Fal'ko, V. I. Monolayer $MoS_2$: Trigonal warping, the Γ valley, and spin-orbit coupling effects. *Physical Review B* **88**, 045416 (2013).

30  Butov, L. V., Lai, C. W., Ivanov, A. L., Gossard, A. C. & Chemla, D. S. Towards Bose-Einstein condensation of excitons in potential traps. *Nature* **417**, 47-52 (2002).
31  Lee, C.-H., Lee, G.-H., Arend M. van der Zande, Chen, W., Li, Y., Han, M., Cui, X., Arefe, G., Nuckolls, C., Tony F. Heinz, Guo, J., James Hone & Kim, P. Atomically thin p-n junctions with van der Waals heterointerfaces. *arXiv:1403.3062* (2014).
32  Furchi, M. M., Pospischil, A., Libisch, F., Burgdörfer, J. & Mueller, T. Photovoltaic effect in an electrically tunable van der Waals heterojunction. *arXiv:1403.2652* (2014).
33  Cheng, R., Li, D., Zhou, H., Wang, C., Yin, A., Jiang, S., Liu, Y., Chen, Y., Huang, Y. & Duan, X. Electroluminescence and photocurrent generation from atomically sharp WSe2/MoS2 heterojunction p-n diodes. *arXiv:1403.3447* (2014).
34  Fang, H., Battaglia, C., Carraro, C., Nemsak, S., Ozdol, B., Kang, J. S., Bechtel, H. A., Desai, S. B., Kronast, F., Unal, A. A., Conti, G., Conlon, C., Palsson, G. K., Martin, M. C., Minor, A. M., Fadley, C. S., Yablonovitch, E., Maboudian, R. & Javey, A. Strong interlayer coupling in van der Waals heterostructures built from single-layer chalcogenides. *arXiv:1403.3754 [*(2014).

Figure 1.

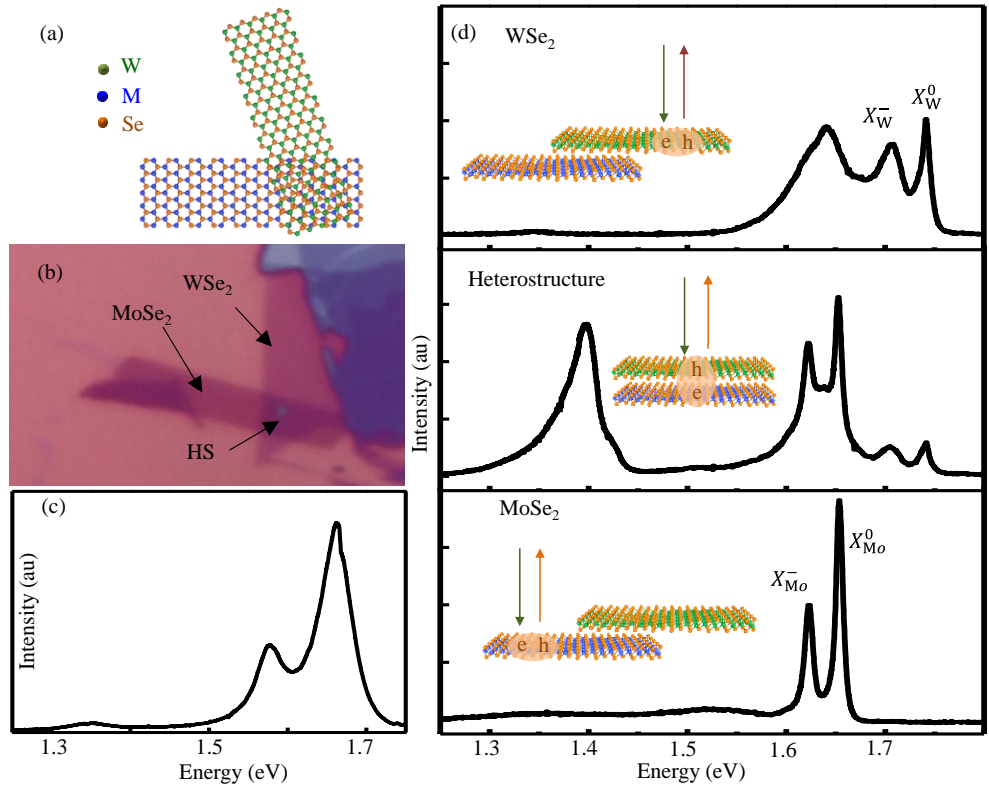

**Figure 1 | Intralayer and interlayer excitons of a monolayer MoSe$_2$-WSe$_2$ vertical heterostructure. a,** Cartoon depiction of a MoSe$_2$-WSe$_2$ heterostructure (HS). **b,** Microscope image of a MoSe$_2$-WSe$_2$ heterostructure (Device 1). **c,** Room temperature photoluminescence of the heterostructure under 20 µW laser excitation at 2.33 eV. **d,** Photoluminescence of individual monolayers and the heterostructure at 20 K under 20 µW excitation at 1.88 eV (plotted on the same scale).

Figure 2.

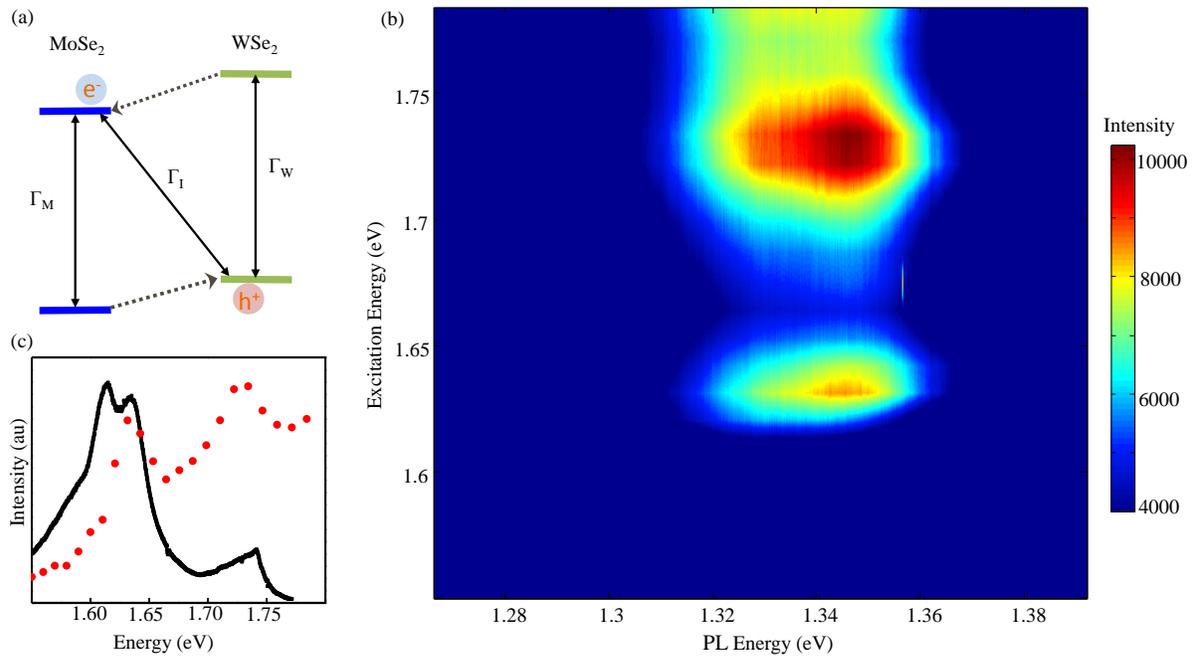

**Figure 2 | Photoluminescence excitation spectroscopy of the interlayer exciton at 20K. a,** Type II semiconductor band alignment diagram for the 2D MoSe$_2$-WSe$_2$ heterojunction. $\Gamma_i$ represents the relaxation rate of the i transition. **b,** PLE intensity plot of the heterostructure region with an excitation power of 30 µW and 5 second CCD integration time. **c,** Spectrally integrated PLE response (red) overlaid on PL (black) with 100 µW excitation at 1.88 eV.

Figure 3.

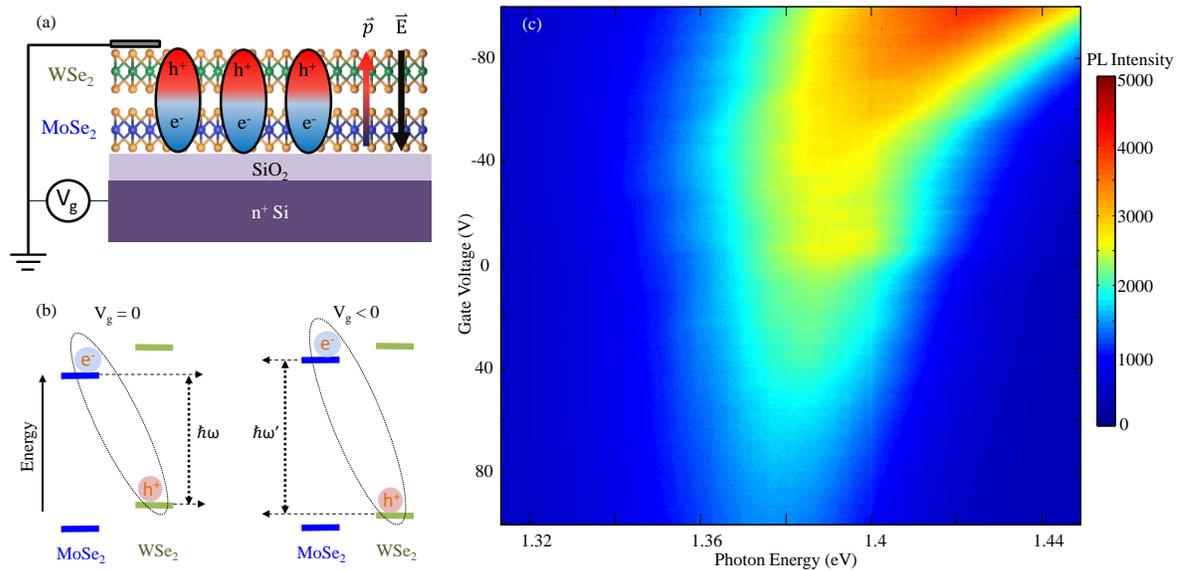

**Figure 3 | Gate control of the interlayer exciton and band-alignment. a,** Device 2 geometry. The interlayer exciton has a permanent dipole, corresponding to an out-of-plane electric polarization. **b,** Electrostatic control of the band alignment and the interlayer exciton resonance. **c,** Colormap of interlayer exciton photoluminescence as a function of applied gate voltage under 70 μW excitation at 1.744 eV, 1 second integration time.

Figure 4.

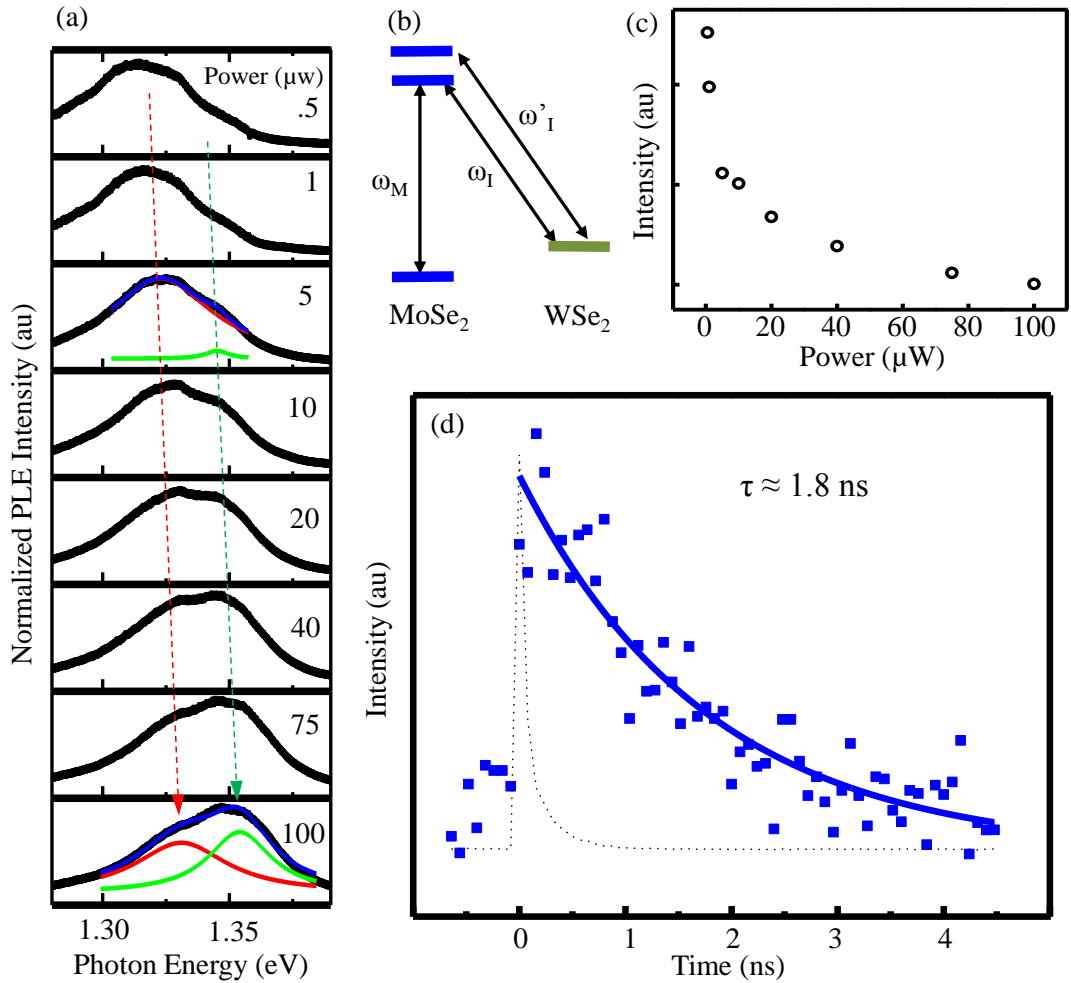

**Figure 4 | Power saturation, MoSe$_2$ conduction band splitting, and the interlayer exciton lifetime at 20K. a,** Power dependence of the interlayer exciton for 1.722 eV excitation with a bi-Lorentzian fit to the 5 and 100 µW plots, normalized for power and CCD integration time. **b,** Illustration of the heterojunction band diagram, including the conduction band splitting of MoSe$_2$. **c,** Spectrally integrated intensity of the interlayer exciton emission, normalized to laser power and integration time, as a function of excitation power shows the saturation effect. **d,** Time resolved PL of the interlayer exciton (1.35 eV) shows a lifetime of about 1.8 ns. The dashed curve is the instrument response to the excitation laser pulse.

# Supplementary Information for

# Observation of Long-Lived Interlayer Excitons in Monolayer MoSe$_2$-WSe$_2$ Heterostructures

**Authors**: Pasqual Rivera[1], John R. Schaibley[1], Aaron M. Jones[1], Jason S. Ross[2], Sanfeng Wu[1], Grant Aivazian[1], Philip Klement[1], Nirmal J. Ghimire[3,4], Jiaqiang Yan[4,5], D. G. Mandrus[3,4,5], Wang Yao[6], Xiaodong Xu[1,2*]

**Supplementary Figures**

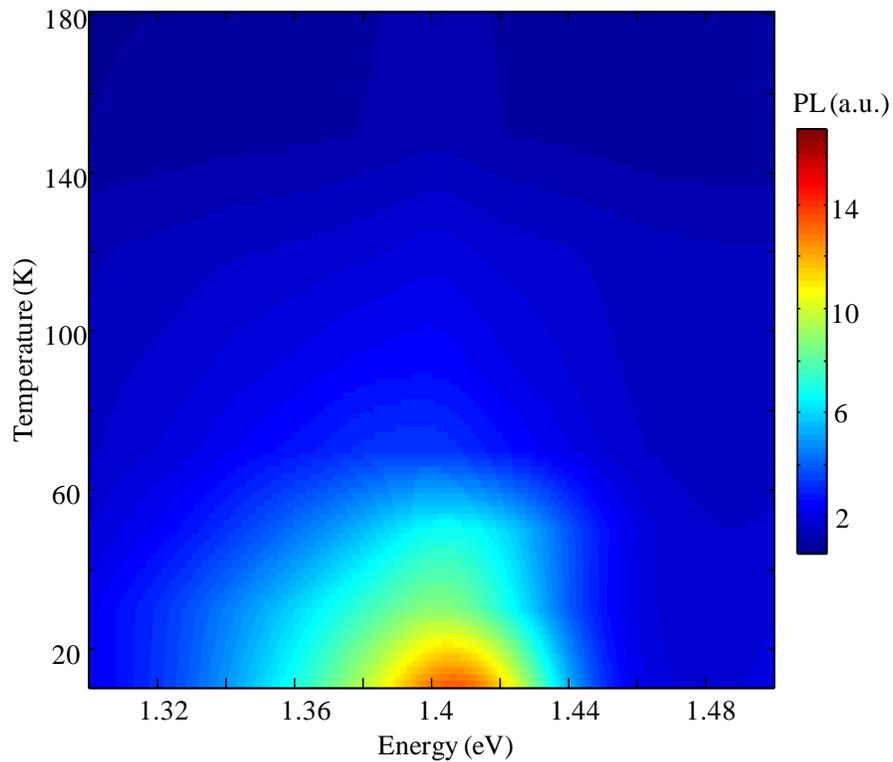

**Figure S1 | Temperature dependence of interlayer exciton.** Photoluminescence (PL) intensity plot showing decreased PL intensity of the interlayer exciton with increasing temperatures.

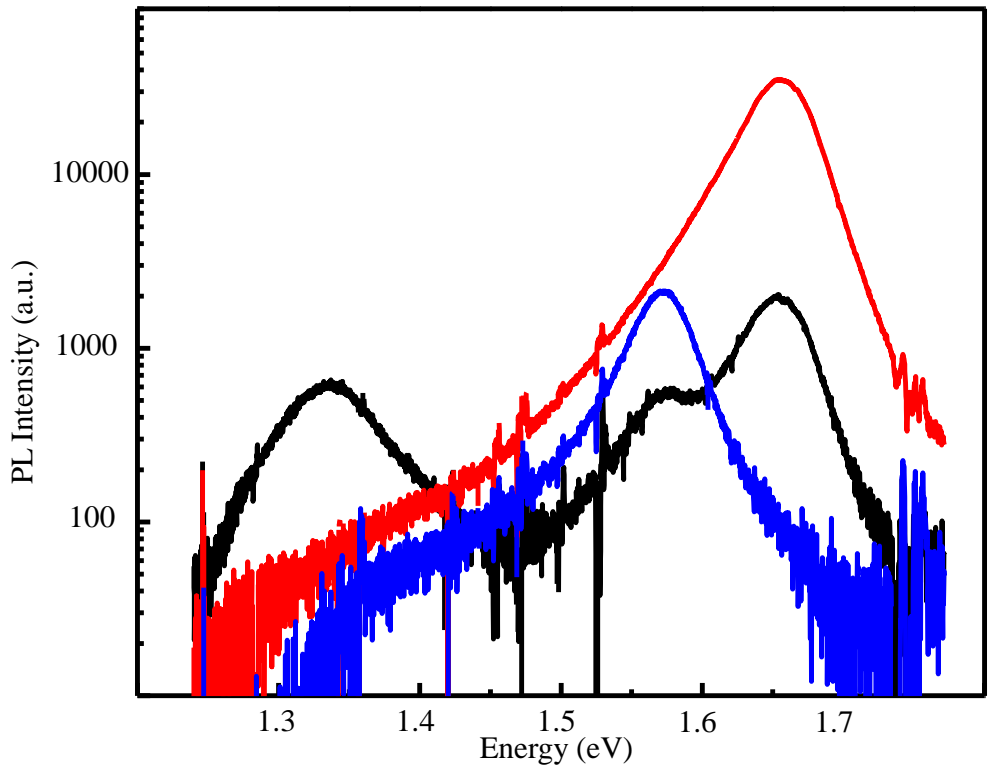

**Figure S2 | Room temperature photoluminescence.** Photoluminescence from individual monolayers of WSe$_2$ (red) and MoSe$_2$ (blue) are quenched by an order of magnitude on the heterostructure (black). Data are taken from device 1 after a thermal cycle.

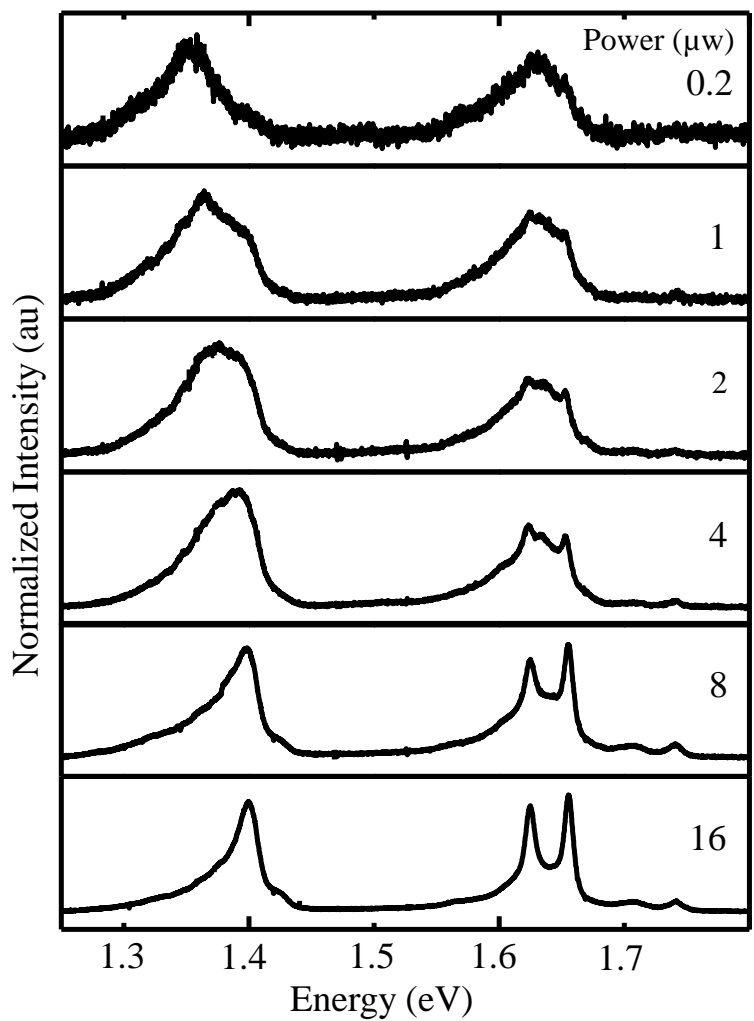

**Figure S3 | Power dependence of interlayer exciton photoluminescence.** Photoluminescence intensity as a function of power for the heterostructure region of device 1, normalized to the peak intensity. Experimental temperature is 20 K and the excitation energy is 1.88 eV. The interlayer peak is composed of a doublet corresponding to the $MoSe_2$ conduction band splitting.